\input{aipcheck}

\documentclass[
    ,final            % use final for the camera ready runs
  ]
  {aipproc}

\layoutstyle{6x9}

\begin{document}

\title{Artificial  variability in XMM-Newton observations of X-ray sources: M31 as a case study}

\classification{95.55.Ka, 95.75.Wx, 95.85.-e, 97.80.Jp, 97.60.Lf, 97.60.Jd, 98.56.Ne, 98.62.Mw}
\keywords      {Xrays: binaries - Galaxies: individual: M31}

\author{R. Barnard}{
  address={The Open University, Walton Hall, Milton Keynes, MK7 6AA, UK}
}
\author{S. Trudolyubov}{
 address={Institute of Geophysics and Planetary Physics, University of California, Riverside, CA 92521, USA}
}
\author{C. A. Haswell}{
address={The Open University, Walton Hall, Milton Keynes, MK7 6AA, UK}
}
\author{U. C. Kolb}{
address={The Open University, Walton Hall, Milton Keynes, MK7 6AA, UK}
}
\author{J. P. Osborne}{
address={The University of Leicester, Leicester, LE1 7RH, UK}
}
\author{ W. H. Priedhorsky}{
address={Los Alamos National Laboratory, Los Alamos, NM 87545}
}

\begin{abstract}
Power density spectra (PDS) that are characteristic of low mass X-ray binaries (LMXBs) have been previously reported for M31 X-ray sources observed by XMM-Newton. However, we have recently discovered that these PDS are false positives resulting from the improper manipulation of non-simultaneous lightcurves. The lightcurves produced by the XMM-Newton Science Analysis Software (SAS)  are non-synchronised by default. This affects not only the combination of lightcurves from the three EPIC detectors (MOS1, MOS2 and pn), but also background subtraction in the same CCD. It is therefore imperative that all SAS-generated lightcurves are synchronised by time filtering, even if the whole observation is to be used. We combined simulated lightcurves at various intensities with various offsets and found that the artefact is more dependent on the offset than the intensity.  While previous timing results from M31 have been proven wrong, and also the broken power law PDS in NGC 4559 ULX-7, XMM-Newton was able to detect aperiodic variability in just 3 ks of observations of NGC 5408 ULX1. Hence XMM-Newton remains a viable tool for analysing variability in extra-galactic X-ray sources. 

\end{abstract}

\maketitle

%%%%%%%%%%%%%%%%%%%%%%%%%%%%%%%%%%%%%%%%%%%%
%% MAINMATTER
%%%%%%%%%%%%%%%%%%%%%%%%%%%%%%%%%%%%%%%%%%%%

\section{Introduction}

The variability and spectral properties of Galactic low mass X-ray binaries (LMXBs) are well known to depend more on the accretion rate than the nature of the primary \citep[neutron star or black hole,][]{vdk94}. At low accretion rates, the power density spectra (PDS) may be characterised by broken power laws, with the spectral index, $\gamma$, changing from $\sim$0 to $\sim$1 at some break frequency in the range 0.01--1 Hz \citep{vdk94}; such variability has a r.m.s. power of $\sim$10--40\% \citep[e.g.][]{vdk95}.  We describe such PDS as Type A \citep{bko04}. At higher accretion rates, the PDS is described by a simple power law with $\gamma$ $\sim$1 and r.m.s. variability $<$10\% \citep{vdk94,vdk95}. We refer to these PDS as Type B \citep{bko04}.

Type A variability is characteristic of disc-accreting XBs, and any X-ray source that exhibits such variability may be identified as an XB, rather than a fore-ground star or background active galaxy \citep[see e.g.][]{bok03}. Such variability has been reported by Barnard et al. in  XMM-Newton observations of M31 \citep[see e.g.][]{bok03, bko04,will05,bar05}. Type A PDS have also been reported for ultraluminous X-ray sources (ULXs) in NGC\thinspace 4559 \citep{crop04}  and NGC\thinspace 5408 \citep{sor04}. 

%Furthermore, we have evidence that the transition from Type A to Type B occurs at $\sim$10\% of the Eddington limit. Since the Eddington limit is proportional to the mass of the primary, black hole LMXBs would exhibit Type A variability at higher luminosities than neutron star LMXBs; we classify any X-ray source that exhibits Type A variability at a 0.3--10 keV luminosity $>$ 4$\times$10$^{37}$ erg s$^{-1}$ as a black hole candidate \citep{bko04,bar05}. 

\section{XMM-Newton lightcurves} 

 However, we have now discovered an additional,  artificial origin for such stochastic variability, caused by the improper addition of lightcurves from the three EPIC instruments on board XMM-Newton (MOS1, MOS2 and pn), where the lightcurves are not synchronised. This artefact is discussed in detail in \citep{bar06}.

\begin{figure}[!t]
\includegraphics[angle=270,scale=.4]{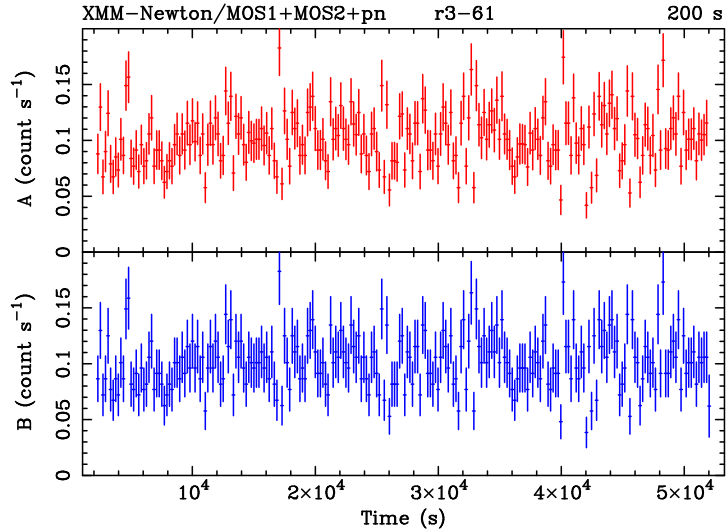}
\caption{ XMM-Newton combined MOS1+MOS2+pn lightcurves of r3-61 in the 0.3--10 keV band; the combination of lightcurves produced by Method I and Method II are shown in the top and bottom panels respectively. The best fit line of constant intensity to the Method I lightcurve gives $\chi^2$/dof = 408/238, while the best fit to the Method II lightcurve gives $\chi^2$/dof = 280/239
}\label{fig1}
\end{figure}

XMM-Newton has observed M31 extensively, and Barnard et al. \citep{bok03, bko03, bko04,bfh06,bar05} have conducted a survey of bright sources in the central region, over four observations. For each observation of every source, 0.3--10 keV  lightcurves were obtained  from source and background regions in the  MOS1, MOS2 and pn images,   using 2.6 second binning.  %X-ray spectra from the pn instrument were also obtained  in the 0.3--10 keV band.

We have recently discovered that the results of manipulating lightcurves produced with SAS tool {\bf evselect} using  {\bf lcmath} produces different results, depending on how the time selection is performed.
The three methods of time selection may be summarised as follows
\begin{itemize}
\item
 {\bf Method I}:~~using the expression ``(TIME in [{\it tstart}:{\it tstop}])'' in {\bf evselect}
\item 
 {\bf Method II}:~~adding the keywords TLMIN1 = {\it tstart} and TLMAX1 = {\it tstop} to the events files for each detector, using the {\bf fparkey} FTOOL
\item
 {\bf Method III}:~~filtering each lightcurve with the {\bf evselect} keywords ``timemin ={\it tstart} timemax={\it tstop}'
\end{itemize}
  Methods II and III are equivalent. Vitally, they are not equivalent to Method I, although this is not mentioned in any documentation.

In Fig.~\ref{fig1}, we compare  MOS1+MOS2+pn lightcurves of the M31 X-ray source XMMU\thinspace J004207.6+411815 (r3-61) from the  2002, January 6 XMM-Newton observation, extracted using Methods I (top) and II (bottom). The best fit line of constant intensity to the Method I lightcurve yields $\chi^2$/dof=408/239, while the best fit to the Method II lightcurve  gives $\chi^2$/dof=280/240. The best fit intensity is 0.102 count s$^{-1}$ for both lightcurves. Consequently, we analysed each lightcurve to determine which, if any, was correct.

We first examined the MOS1, MOS2 and pn source lightcurves individually before background subtraction, testing whether the errors were simply Gaussian. We binned each lightcurve to 100 s and obtained the count rate and error for each bin of the resulting lightcurve. 
We found that each of the input lightcurves had been assigned Gaussian uncertainties (uncertainty $\sigma$ = $N^{0.5}$ for $N$ counts in a time bin), so the correct combined lightcurve would have Gaussian uncertainties also. In fact, Gaussian uncertainties underestimate the confidence limits at low count rates \citep{gehr86}.

\begin{figure}[!t]
\includegraphics[angle=270,scale=.5]{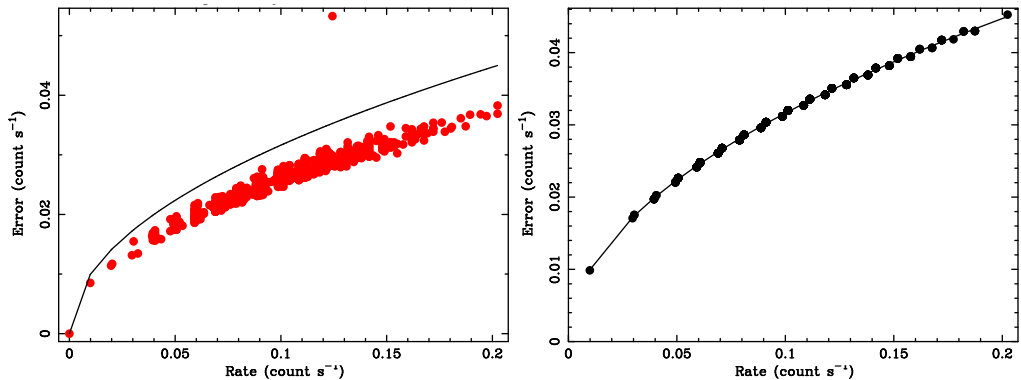}
\caption{ Error vs. intensity for data points in MOS1+MOS2+pn lightcurves of r3-61 combined using Method I (left) and Method II (right). In each case, a curve represents Gaussian errors. Clearly, the lightcurves produced using Method I were incorrectly added, as the errors are mostly smaller than Gaussian; meanwhile the Method II lightcurve has Gaussian errors, showing that it has been correctly treated.  
}\label{fig3}
\end{figure}

The rate vs. errors for the combined MOS1+MOS2+pn lightcurves using Method I and Method II are presented in the left and right panels respectively of Fig~\ref{fig3}. It is clear that the Method II lightcurve has been correctly produced, as the errors are Gaussian; however, the errors for the Method I lightcurve are unfeasibly small, as they are smaller than the Gaussian limit; as a result, the Method I lightcurve is artificially variable.

Closer inspection of the lightcurves has revealed the difference between Method I and Method II lightcurves: the Method II lightcurves are synchronised, while the Method I lightcurves are not.  {\bf Lcmath} processed the latter  incorrectly, resulting in  the artificial variability.  

\begin{figure}[!t]
\includegraphics[angle=270,scale=.5]{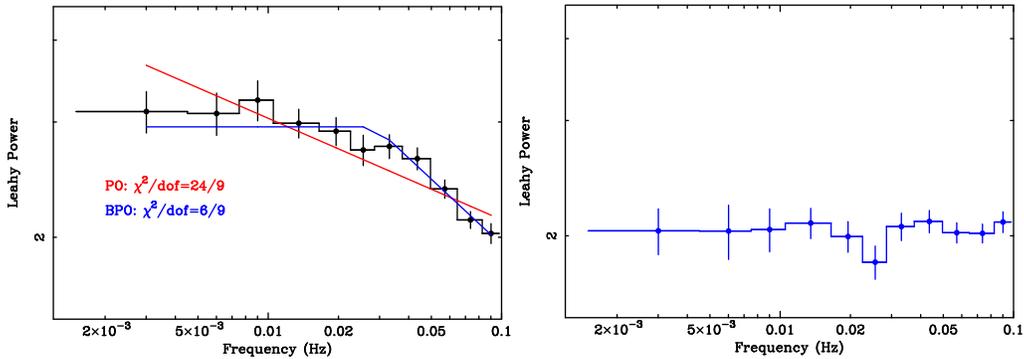}
\caption{ Power density spectra of 0.3--10 keV lightcurves of r3-60, using Method I (left) and Method II (right). We present best fits to the Method I PDS of power law (PO) and broken power law models (BPO); a broken power law is clearly required. The PDS is hence reminiscent of Galactic LMXBs. However, the Method II PDS is flat, revealing that this variability is artificial.
}\label{fig4}
\end{figure}

Many PDS from combined lightcurves obtained using Method I exhibit Type A variability, while those obtained with Method II do not.  In order to investigate the cause of the Type A variability in the observed PDS, we examined the Method I and II lightcurves of XMMU\thinspace J004208.9+412048 (r3-60). The Method I and Method II PDS are shown in the left and right panels respectively of Fig.~\ref{fig4}; they were obtained using the FTOOL {\bf powspec}.  The Method I PDS is well modelled by a broken power law ($\chi^2$/dof = 6/8). Such a PDS is characteristic of Galactic LMXBs. However, the PDS of the Method II lightcurve is consistent with Poisson noise; hence the variability in the  Method I PDS is artificial.

Following this work, we produced background-subtracted EPIC-pn lightcurves of various X-ray sources from the 2002 January XMM-Newton observation of the M31 core, to see if the lightcurves from the same detector was synchronised.  The Method I PDS again is of Type A. Hence,  background subtraction also results in artefacts when the lightcurves are not synchronised.

We have investigated the artefact by adding simulated lightcurves with various intensities and different delays between lightcurves \citep{bar06}. We find that the artefact is determined by the delays, but independent of the lightcurve intensities. Hence it is vitally important to synchronise all XMM-Newton observations, even if the whole observation is to be used. 

\section{Ultra-luminous X-ray sources in NGC 4559 and NGC~5408}
\label{ulx}
%\begin{figure}[!t]
%\resizebox{\hsize}{!}{\includegraphics[angle=270]{combpds.eps}}
%\caption{ PDS of Method I lightcurves of r3-60 from pn (a), MOS1 (b), MOS2 (c) are compared with the combined EPIC Method I PDS (d). 
%}\label{fig4a}
%\end{figure}

 We found published broken power law PDS in NGC 4559 ULX7 \citep{crop04} and also in a ULX in NGC 5408 \citep{sor04}. We re-analysed XMM-Newton observations of these sources, first following the methods described in these papers, using non-synchronised lightcurves, then comparing their PDS with those from synchronised lightcurves. 

\begin{figure}[!t]
\includegraphics[angle=270,scale=.5]{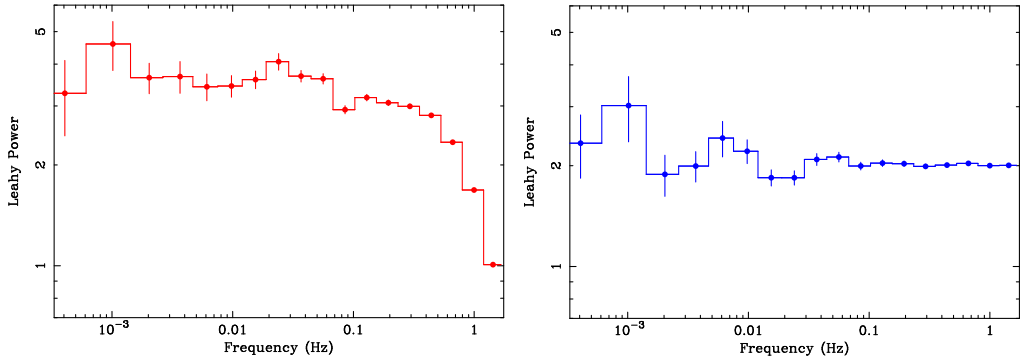}
\caption{ Power density spectra of 0.2--12 keV lightcurves of the NGC 4559 ULX7 using Method I (left) and Method II (right). The PDS are Leahy normalised, so that Poisson noise has a power of 2. While a broken power law PDS is observed in the Method I PDS, there is no excess power observed in the Method II PDS. Hence the broken power law PDS reported for NGC 4559 ULX7 is artificial. 
}\label{fig6a}
\end{figure}

Cropper et al. \citep{crop04} reported a broken power from an  ultra-luminous X-ray source (ULX) in NGC 4559 with a UV/X-ray luminosity exceeding 2.1$\times$10$^{40}$ erg s$^{-1}$. Following their example, we extracted MOS1, MOS2 and pn lightcurves from a source region with 30$''$ radius using Methods I and II. The resulting PDS are shown in the left and right panels of Fig.~\ref{fig6a} respectively. The Method I PDS exhibits Type A variability, while the Method II PDS does not, revealing that the PDS reported for ULX 7 by \citep{crop04} is artificial.

\begin{figure}[!t]
\includegraphics[angle=270,scale=.3]{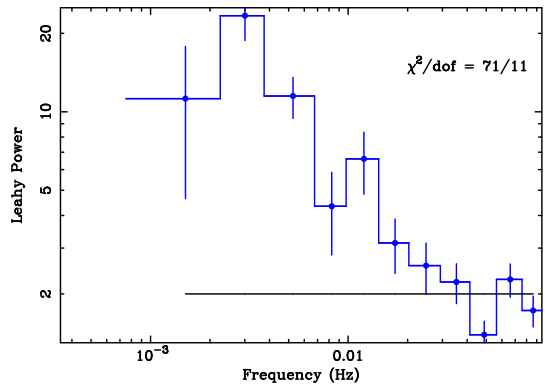}
\caption{Method II PDS for the NGC 5408 ULX, averaged over 9 intervals of 64 bins. The PDS is Leahy normalised, and a line of constant power equivalent to the Poisson noise was fitted; the resulting $\chi^2/dof$ = 71/11, showing that the source is variable.
}\label{fig8}
\end{figure}

Soria et al. \citep{sor04} studied a series of XMM-Newton observations of a ULX in the dwarf galaxy NGC 5408. 
A Method II PDS is presented in Fig.~\ref{fig8}; fitting this PDS with a Leahy Power of 2 (Poisson noise) results in a $\chi^2/dof$ of 71/11, hence the variability is significant.
 However, there is no statistical requirement for the claimed break in the PDS in the $\sim$3 ks of good data; a longer observation would be required to identify any break.

%\begin{figure}[!t]
%\resizebox{\hsize}{!}{\includegraphics[angle=270]{ngc5408nspds.eps}}
%\caption{ Power density spectra of 0.2--12 keV lightcurves of the NGC 5408 ULX using method A (left) and Method II (right). The y axis is scaled to the fractional r.m.s. variability, in keeping with \citepp{sor04}; this scaling is simply Leahy power / $\bar{x}$. Both PDS appear to be broken power laws, indicating that the variability is intrinsic to the source.
%}\label{fig7}
%\end{figure}

\section{Conclusions}
\label{conc}
Some of the authors have previously reported variability in XMM-Newton observations of M31 X-ray sources that are characterised by broken power law PDS, which are signatures of disc-accreting X-ray binaries \citep{bok03,bko04,will05,bar05}. However, these ``Type A'' PDS observed in XMM-Newton observations of M31 were found to be false positives, and may be entirely attributed to errors introduced by {\bf lcmath} when combining non-synchronised lightcurves. It is therefore imperative to synchronise all lightcurves; this is most easily achieved by following Method II.

 Extragalactic analogues of the Galactic X-ray binaries should exhibit Type A and Type B behaviour; Type A variability should be observed in sufficiently bright X-ray sources. %In addition, we still expect X-ray sources that exhibit Type A variability $>$4$\times$10$^{37}$ erg s$^{-1}$ in the 0.3--10 keV band to be black hole LMXBs.
We have shown the broken power law PDS of NGC 4559 ULX7 to be artificial also; however, the PDS of the NGC 5408 ULX shows that timing analysis of XMM-Newton observations of extra-galactic X-ray sources is viable  after all.  Indeed, M31 X-ray sources have exhibited  pulsations \citep{osb01,tru05}; bursts \citep{ph05}; periodic dipping due to photoelectric absorption \citep{tru02,man04}, from a precessing disc in one case \citep{bfh06}; lastly, branch movement in a trimodal colour-intensity diagram reminiscent of Galactic Z-sources \citep{bko03}.

\begin{theacknowledgments}
Power density spectra were fitted using {\bf fitpowspec}, provided by P.J. Humphrey. R.B is funded by PPARC, S.T is supported by NASA grant  NAG5-12390.
\end{theacknowledgments}

%%%%%%%%%%%%%%%%%%%%%%%%%%%%%%%%%%%%%%%%%%%%%%%%
%% The bibliography can be prepared using the BibTeX program or
%% manually.
%%
%% The code below assumes that BibTeX is used.  If the bibliography is
%% produced without BibTeX comment out the following lines and see the
%% aipguide.pdf for further information.
%%
%% For your convenience a manually coded example is appended
%% after the \end{document}
%%%%%%%%%%%%%%%%%%%%%%%%%%%%%%%%%%%%%%%%%%%%%%%%

%%%%%%%%%%%%%%%%%%%%%%%%%%%%%%%%%%%%%%%%%%%%%%%%
%% You may have to change the BibTeX style below, depending on your
%% setup or preferences.
%%
%%
%% For The AIP proceedings layouts use either
%%%%%%%%%%%%%%%%%%%%%%%%%%%%%%%%%%%%%%%%%%%%

\bibliographystyle{aipproc}   % if natbib is available
%\bibliographystyle{aipprocl} % if natbib is missing

%%%%%%%%%%%%%%%%%%%%%%%%%%%%%%%%%%%%%%%%%%%
%% You probably want to use your own bibtex database here
%%%%%%%%%%%%%%%%%%%%%%%%%%%%%%%%%%%%%%%%%%%
\bibliography{m31}

%%%%%%%%%%%%%%%%%%%%%%%%%%%%%%%%%%%%%%%%%%%
%% Just a reminder that you may have to run bibtex
%% All of it up to \end{document} can be removed
%% if you don't like the warning.
%%%%%%%%%%%%%%%%%%%%%%%%%%%%%%%%%%%%%%%%%%%
\IfFileExists{\jobname.bbl}{}
 {\typeout{}
  \typeout{******************************************}
  \typeout{** Please run "bibtex \jobname" to optain}
  \typeout{** the bibliography and then re-run LaTeX}
  \typeout{** twice to fix the references!}
  \typeout{******************************************}
  \typeout{}
 }

\end{document}